\documentclass[preprint,aps,floatfix,nofootinbib,groupedaddress]{revtex4}
\usepackage{float}
\usepackage{amsfonts}
\usepackage{amssymb}
\usepackage{multirow}
\usepackage{xparse}
\usepackage{savesym}
\usepackage{amsmath}
\savesymbol{iint}
\usepackage{txfonts}
\restoresymbol{TXF}{iint}
\usepackage{ctable}

\usepackage{epsfig}
\usepackage{textcomp}
\usepackage[page]{appendix}
\usepackage{bm}

\begin{document}

\title{Merged-Beams for Slow Molecular Collision Experiments}

\author{Qi Wei, Igor Lyuksyutov and Dudley Herschbach\footnote{ Corresponding email: dherschbach@gmail.com}}
\address{Department of Physics, Texas A$\&$M University, College Station, TX 77843, USA}

\begin{abstract}
Molecular collisions can be studied at very low relative kinetic
energies, in the milliKelvin range, by merging codirectional beams
with much higher translational energies, extending even to the
kiloKelvin range, provided that the beam speeds can be closely
matched.  This technique provides far more intensity and wider
chemical scope than methods that require slowing both collision
partners. Previously, at far higher energies, merged beams have been
widely used with ions and/or neutrals formed by charge transfer.
Here we assess for neutral, thermal molecular beams the range and
resolution of collision energy that now appears attainable,
determined chiefly by velocity spreads within the merged beams.  Our
treatment deals both with velocity distributions familiar for
molecular beams formed by effusion or supersonic expansion, and an
unorthodox variant produced by a rotating supersonic source capable
of scanning the lab beam velocity over a wide range.
\end{abstract}

\maketitle

\section{Introduction}

The frontier field of cold ($< 1$ K) and ultracold ($< 1$ mK)
gas-phase molecular physics has brought forth many innovations
\cite{Book1,Carr,Faraday,Dulieu}. Among motivating challenges is the
prospect of studying collision processes, especially chemical
reactions, under "matterwave" conditions. Reaching that realm, where
quantum phenomena become much more prominent than in ordinary "warm"
collisions, requires attaining {\it relative} velocities so low that
the deBroglie wavelength becomes comparable to or longer than the
size of the collision partners. That has been achieved recently for
reactions of alkali atoms with alkali dimers formed from ultracold
trapped alkali atoms by photoassociation or Feshbach resonances
\cite{Ospelkaus,Miranda}. With the aim of widening the chemical
scope, much effort has been devoted to developing means to slow and
cool preexisting molecules. (Compilations are given in
\cite{Meerakker,Bell,Hogan,Sheffield}.) For chemical reactions,
however, as yet it has not proved feasible to obtain sufficient
yields at very low collision energies, using either trapped
reactants or crossed molecular beams. The major handicap in such
experiments is that {\it both} reactants must contribute adequate
flux with very low translational energy.

Merged codirectional beams with closely matched velocities offer a
way to obtain far higher intensity at very low {\it relative}
collision energies, since then {\it neither} reactant needs to be
particularly slow. Moreover, many molecular species not amenable for
slowing techniques become available as reactants.  Merged beams have
been extensively used with ions and/or neutrals formed by charge
transfer, to perform experiments at relative energies below 1 eV
with beams having keV energies \cite{Phaneuf}. A key advantage is a
kinematic feature that deamplifies contributions to the relative
energy by velocity spreads in the parent beams. By virtue of precise
control feasible with ions, the velocity spreads are also quite
small, typically $\sim$0.1\% or less. For thermal molecular beams,
such as we consider here, the spreads are usually $\sim$10\% or
more. That enables attaining low relative collision energy, but much
lower energies and improved resolution can be obtained by narrowing
the spreads to $\sim$1\%, which now appears feasible at an
acceptable cost in intensity. Surprisingly, application of merged
beams to low-energy collisions of neutral molecules has been long
neglected. We have come across only three previous, very brief
suggestions \cite{Book2000,Gupta,Meeakker2}. Our treatment
accompanies experiments now underway at Texas A\&M University
\cite{Sheffield}.

In Sec. II, in order to assess the major role of velocity spreads in
merged beams, we evaluate the average relative kinetic energy,
$\langle E_R\rangle$ and its rms spread $\triangle E_R$ by
integrating over velocity distributions familiar for molecular
beams.  Reduced variable plots are provided that display the
dependence on the ratio of most probable velocities in the merged
beams, their velocity spreads, and merging angle.  In Secs. III and
IV we discuss experimental prospects, limitations, and options.

\section{Averages over Velocity Distributions}

For beams with lab speeds $V_1$ and $V_2$ intersecting at an angle
$\theta$, the relative kinetic energy is
\begin{equation}
E_R = \frac{1}{2}\mu\left(V_1^2+V_2^2-2V_1V_2\text{cos}\theta\right)
\end{equation}
with $\mu = m_1m_2/(m_1 + m_2)$ the reduced mass. For merged beams
it is feasible to restrict the angle $\theta$ to a small spread,
fixed by geometry and typically only about a degree or so about
$\theta = 0^\circ$. Here we evaluate the average of $E_R$ over the
beam velocity distributions,
\begin{equation}
\langle E_R \rangle = \frac{1}{2}\mu\left[\langle V_1^2\rangle +
\langle V_2^2\rangle - 2\langle V_1\rangle \langle
V_2\rangle\text{cos}\theta\right]
\end{equation}
and the rms spread,
\begin{equation}
\triangle E_R = \left[\langle E_R^2\rangle - \langle
E_R\rangle^2\right]^{1/2}.
\end{equation}
These require only $\langle V^k\rangle$ , with k = 1 - 4, for the
individual beams. We obtain analytic expressions for averages over
velocity distributions for beams formed by effusive flow, by
supersonic expansion, and by a rotating supersonic nozzle. Figure 1
illustrates these distributions. For the supersonic beams, three
widths are shown; the broadest ($\sim$ 10\%) is typical, the
narrowest ($\sim$ 1\%) is near the best achieved in Stark or Zeeman
molecular decelerators exploiting phase stability and transverse
focusing \cite{Meerakker,Hogan}. Results given here pertain to
molecular flux distributions; as noted in an Appendix, they are
readily adapted for number density distributions.

{\bf A. Effusive beams.}

The flux distribution,
\begin{equation}
F(V) = V^3\text{exp}[-(V/\alpha)^2]
\end{equation}
is governed by a single parameter, $\alpha = (2k_BT_0/m)^{1/2}$,
with $k_B$ the Boltzmann constant, and $T_0$ the source temperature.
The averaged powers of the velocity are
\begin{equation}
\langle (V/\alpha)^k\rangle = \int_0^\infty t^kF(t)dt/\int_0^\infty
F(t)dt = \Gamma[(k+4)/2]
\end{equation}
with $t = V/\alpha$ and $\Gamma(z)$ the Gamma function. The most
probable velocity, $V_{mp}/\alpha = (3/2)^{1/2} = 1.224$, and the
rms velocity spread is $(\triangle V/\alpha)_{rms} = [2 -
(\tfrac{3}{4}\pi^{1/2})^2]^{1/2} = 0.483$. The average beam kinetic
energy is
\begin{equation}
\langle E_{BK}\rangle/(\tfrac{1}{2}m \alpha^2) = \langle
(V/\alpha)^2\rangle = 2
\end{equation}
and the rms kinetic energy spread is
\begin{equation}
\triangle E_{rms}/(\tfrac{1}{2}m\alpha^2) = [\langle
(V/\alpha)^4\rangle - \langle (V/\alpha)^2\rangle^2]^{1/2} =
[\Gamma(4)-\Gamma^2(3)]^{1/2} = [6-4]^{1/2} = 1.41
\end{equation}
The spread thus is comparable to the beam kinetic energy.

For a merged pair of effusive beams, it is convenient to define
$\alpha_1 = \alpha_{12}\text{cos}\phi$, $\alpha_2 =
\alpha_{12}\text{sin}\phi$, with $\alpha_{12} = (\alpha_1^2 +
\alpha_2^2)^{1/2}$, and $\langle v^k\rangle = \langle
(V_1/\alpha_1)^k\rangle = \langle (V_2/\alpha_2)^k\rangle =
\Gamma[(k+4)/2]$. The relative kinetic energy is
\begin{equation}
\langle E_R\rangle/(\tfrac{1}{2}\mu\alpha_{12}^2) = \langle
v^2\rangle - 2\text{cs}\langle v\rangle^2\text{cos}\theta
\end{equation}
with c = cos$\phi$ and s = sin$\phi$. The rms spread is
\begin{equation}
\triangle E_R/(\tfrac{1}{2}\mu\alpha_{12}^2) = \left[ A -
B\text{cos}\theta + C\text{cos}^2\theta\right]^{1/2}
\end{equation}
with
\begin{equation*}
A=(c^2+s^2)\left(\langle v^4\rangle-\langle v^2\rangle^2\right)
\end{equation*}
\begin{equation*}
B=4cs\left(\langle v\rangle \langle v^3\rangle-\langle
v\rangle^2\langle v^2\rangle\right)
\end{equation*}
\begin{equation*}
C=4c^2s^2\left(\langle v^2\rangle^2-\langle v\rangle^4\right)
\end{equation*}
Figure 2 shows how $\langle E_R\rangle$ and $\triangle E_R$ vary
with the ratio of beam velocities, which is proportional to
$\alpha_2/\alpha_1 = \text{tan}\phi$. The curves given are for
intersection angles near zero, pertinent for merged beams. For
matched beam velocities, with $\alpha_2 = \alpha_1$, both $\langle
E_R\rangle$ and $\triangle E_R$ are smallest. There, for small
$\theta$, $\langle E_R\rangle$ and $\triangle E_R$ are less than the
nominal relative kinetic energy, $E_R^\circ =
\tfrac{1}{2}\mu(\alpha_1^2 + \alpha_2^2)$, for a perpendicular
collision ($\theta = 90^\circ$) by factors of only about 5 and 3,
respectively. As $\alpha_2$ decreases below $\alpha_1$, $\langle
E_R\rangle$ and $\triangle E_R$ increase in nearly parallel fashion.
Accordingly, whether or not the merged beam velocities are closely
matched, $\triangle E_R$ exceeds $\langle E_R\rangle$ appreciably
(by $\sim$ 40\% when $\alpha_2 = \alpha_1$). The resolution of the
relative kinetic energy hence is worse than seen in Eqs. (6) and (7)
for the beam kinetic energy itself.

The velocity distribution for effusive beams is the same as for a
bulk gas.  Thus, $\langle E_R\rangle$ and $\triangle E_R$ for gas
mixtures cooled by cryogenic means can be obtained from the merged
beam results by merely setting $\text{cos}\theta  = 0$, equivalent
to integrating over all angles of collision.

{\bf B. Stationary supersonic beams.}

A standard approximation,
\begin{equation}
F(V) = V^3\text{exp}\{ -[(V-u)/\triangle v]^2 \}
\end{equation}
for supersonic beams characterizes the velocity distribution by the
flow velocity u along the centerline of the beam, and a width
parameter $\triangle v = \alpha_{\|} = (2k_BT_{\|}/m)^{1/2}$, where
$T_{\|}$, termed the parallel or longitudinal temperature, pertains
to the molecular translational motion relative to the flow velocity
\cite{Book1988}. According to the thermal conduction model
\cite{Klots}, $T_{\|}/T_0$ is determined by the pressure within the
source, $P_0$, the nozzle diameter, $d$, and the heat capacity
ratio, $\gamma = C_P/C_V$. Likewise, the flow velocity is given by
\begin{equation}
u = (2k_BT_0/m)^{1/2}[\gamma/(\gamma - 1)]^{1/2}[1 -
(T_{\|}/T_0)]^{1/2}
\end{equation}

Analytic results for the velocity averages are readily obtained,
\begin{equation}
\langle (V/u)^k\rangle = P_{k+3}(x)/P_3(x)
\end{equation}
with $x = \triangle v/u$ the ratio of velocity width to flow
velocity, and
\begin{equation}
P_n(x) =  \int_0^\infty t^n\text{exp}\{-[(t-1)/x]^2\}dt
\end{equation}
with $t=V/u$. In the Appendix we give exact analytic formulas for
the $P_n(x)$ functions; Table I provides polynomial approximations;
for $x < 0.3$ these are accurate to better than 0.03\%.

The most probable velocity is $V_{mp} = \tfrac{1}{2}u\{1 + [1 +
4x^2]^{1/2}\} \approx u(1 + x^2)$. The rms velocity spread is
$(\triangle V/u)_{rms} = [P_5/P_3  - (P_4/P_3)^2]^{1/2}$.   The
corresponding average beam kinetic energy is
\begin{equation}
\langle E_{BK}\rangle/(\tfrac{1}{2}mu^2) = P_5/P_3 = \langle (V/u)^2
\rangle
\end{equation}
and the rms spread in kinetic energy is
\begin{equation}
\triangle E_{BK}/(\tfrac{1}{2}mu^2) = \left[P_7/P_3  -
(P_5/P_3)^2\right]^{1/2}
\end{equation}

For a pair of merged supersonic beams, the averaged relative kinetic
energy is
\begin{equation}
\langle E_R\rangle/(\tfrac{1}{2}\mu u_{12}^2) = c^2\langle
v_1^2\rangle + s^2\langle v_2^2\rangle - 2\text{cs}\langle
v_1\rangle\ \langle v_2\rangle\text{cos}\theta
\end{equation}
where $u_{12} = (u_1^2+u_2^2)^{1/2}$. The rms spread is
\begin{equation}
\triangle E_R/\left(\tfrac{1}{2}\mu u_{12}^2\right) = [A -
B\text{cos}\theta + C\text{cos}^2\theta]^{1/2}
\end{equation}
with
\begin{equation*}
A=c^4\left[\langle v_1^4\rangle-\langle v_1^2\rangle^2\right] +
s^4\left[\langle v_2^4\rangle-\langle v_2^2\rangle^2\right]
\end{equation*}
\begin{equation*}
B=4cs\left[c^2\left(\langle v_1^3\rangle - \langle v_1\rangle
\langle v_1^2\rangle\right) \langle v_2\rangle + s^2\left(\langle
v_2^3\rangle - \langle v_2\rangle \langle v_2^2\rangle\right)
\langle v_1\rangle\right]
\end{equation*}
\begin{equation*}
C=4c^2s^2\left[\langle v_1^2\rangle \langle v_2^2\rangle -\langle
v_1\rangle^2 \langle v_2\rangle^2 \right]
\end{equation*}
These expressions are akin to Eqs. (8) and (9), with $\alpha_1$ and
$\alpha_2$ replaced by $u_1$ and $u_2$.   Here $\langle v_i^k\rangle
= \langle(V_i/u_i)^k\rangle$, given by Eq.(12), may differ for the
two beams ($i$ = 1, 2) if their velocity widths ($x_i = \triangle
v_i/u_i$) differ.

Figure 3 displays the dependence of $\langle E_R\rangle$ and
$\triangle E_R$ on the ratio of flow velocities of the beams,
$u_2/u_1 = \text{tan}\phi$, and their velocity widths. Curves are
shown for $\theta = 1^\circ$ and $2^\circ$, to illustrate that the
dependence on the merging angle is weak if the velocity widths of
the beams are fairly large ({\it cf}. Fig.2) but for $\langle
E_R\rangle$ becomes significant if the velocity widths are small
({\it cf}. Fig. 6 below). The dependence on the magnitude of the
flow velocities is included simply by adopting units for $\langle
E_R\rangle$ and $\triangle E_R$ that compare them with the relative
kinetic energy for collisions at about the most probable beam
velocities (nearly equal to $u_1$, $u_2$) at right angles ($\theta=
90^\circ$), or equivalently in a bulk gas.  We designate that by
$E_R^\circ$. As seen in panel (a), if the merged beam velocities are
precisely matched ($u_2 = u_1$; $x_2 = x_1$) the collision kinetic
energy $\langle E_R\rangle$ is very sensitive to the velocity
widths. In contrast, when the beam flow velocities are unmatched by
more than about 15\% (i.e., $u_2/u_1 < 0.85$), the ratio $\langle
E_R\rangle/E_R^\circ$ becomes nearly independent of the velocity
widths and grows larger as the unmatch increases. Panel (b) shows
that regardless of whether the beam velocities are matched or not,
the spread in relative kinetic energy, $\triangle E_R$, varies
strongly with the velocity widths. Panel (c) plots the ratio
$\triangle E_R/\langle E_R\rangle$, which defines the energy
resolution. Whereas for closely matched beams $\langle E_R\rangle$
is minimal, $\triangle E_R$ then approaches its maximal value.
Indeed, for matched beams with $x > 0.05$, the resolution ratio,
$\triangle E_R/\langle E_R\rangle$, is near $2^{1/2}$; that is just
as poor as found in Fig. 2 for effusive beams.

To improve the resolution ratio for matched beams requires narrowing
the velocity widths.  Table II compares, both for the single beam
and matched merged beams, effects of reducing the spread from $x$ =
0.1 to 0.01.  For the single beam, the change in average kinetic
energy is very slight, whereas the rms spread in the kinetic energy
shrinks tenfold.   For the merged beams, the relative kinetic energy
is lowered by a factor of 25, and its rms spread by a factor of 100;
so the resolution ratio is only improved fourfold.  The resolution
ratio can be lowered further if the merged beam velocities are
unmatched, but that raises the averaged relative kinetic energy.
Figure 4 displays the trade-offs involved. To obtain optimally low
$\langle E_R\rangle$ requires nearly exact matching; that can
provide $\langle E_R\rangle/E_R^\circ = 2 \times 10^{-4}$ for $x$ =
0.01 or $3 \times 10^{-4}$ for $x$ = 0.02. However, for exact
matching, the resolution ratio is only $\triangle E_R/\langle
E_R\rangle$ = 0.35 for $x$ = 0.01 and surges to 0.8 for $x$ = 0.02.
To attain resolution of 0.2 or 0.3 even with $x$ = 0.01 requires
$u_2/u_1$ = 0.91 or 0.94 and hence would increase $\langle
E_R\rangle/E_R^\circ$ to $\sim 5 \times 10^{-3}$ or $2 \times
10^{-3}$, respectively. The upshot is, to improve the resolution
ratio by a factor of less than 2 (from 0.35 to 0.2) by unmatching,
requires increasing $\langle E_R\rangle$ by a factor of 25.

\begin{table}[H]
\begin{minipage}{16cm}
\centering \caption{Approximate integrals for velocity
averages.$^a$}
\begin{tabular}{cc}
\hline\hline
$n$\;\;\;\; &\;\;\;\; $A_n(x,y)$ \\
\hline\let\thefootnote\relax\footnotetext{$^a$The $P_n$ functions:
$P_n(x) = P_n(x, 1)$; $P_n(x,y)$; $P_n(x,y,z)$ defined in Eq.(13),
and in Eqs.(A1) and (A3) of the Appendix, respectively, are all well
approximated by $P_n = (\sqrt{\pi}/8)A_n(x, y)$.   For $x < 0.3$ and
$y > 0.5$, the error in this approximation for $P_n$ is $<$ 0.03\%.
}
2\;\;\;\; & \;\;\;\;  $4(2y^2+x^2)x$ \\
3 \;\;\;\;&\;\;\;\;   $4(2y^2+3x^2)xy$ \\
4 \;\;\;\;&\;\;\;\;   $2(4y^4+12x^2y^2+3x^4)x$ \\
5 \;\;\;\;&\;\;\;\;   $2(4y^4+20x^2y^2+15x^4)xy$ \\
6 \;\;\;\;&\;\;\;\;   $(8y^6+60x^2y^4+90x^4y^2+15x^6)x$ \\
7 \;\;\;\;&\;\;\;\;   $(8y^6+84x^2y^4+210x^4y^2+105x^6)xy$  \\
\hline\hline
\end{tabular}
\end{minipage}
\label{table1}
\end{table}

\begin{table}[H]
\begin{minipage}{16cm}
\centering \caption{Comparison of averages and spreads.$^a$}
\begin{tabular}{cccc}
\hline\hline
 Quantity \;\;\;\;&\;\;\;\; Formula    \;\;\;\;&\;\;\;\;\;\;\;  $x = 0.10$   \;\;\;\;&\;\;\;\;   $x = 0.01$  \\
\hline\let\thefootnote\relax\footnotetext{\;\;\;\;\;\;\;\;\;\;\;$^a$For
supersonic beams with $u_2 = u_1$ and $\theta = 1^\circ$; see Eqs.
14-17 and Figs. 3, 4, and 6.}
$V_{mp}/u$                 \;\;\;\;       &\;\;\;\;  $\sim (1+x^2)$                  \;\;\;\;&\;\;\;\;   1.010 \;\;\;\;&\;\;\;\;   1.0001 \\
$\triangle V/u$             \;\;\;\;      &\;\;\;\;  $[(P_5/P_3)-(P_4/P_3)^2]^{1/2}$ \;\;\;\;&\;\;\;\;   0.070 \;\;\;\;&\;\;\;\;   0.0071 \\
$\triangle V/V_{mp}$        \;\;\;\;      &\;\;\;\;                                  \;\;\;\;&\;\;\;\;   0.070 \;\;\;\;&\;\;\;\;   0.0071 \\
$\langle E_{BK}\rangle/(\tfrac{1}{2}mu^2)$ \;\;\;\; &\;\;\;\;  $P_5/P_3$                       \;\;\;\;&\;\;\;\;   1.035 \;\;\;\;&\;\;\;\;   1.0003 \\
$\triangle E_{BK}/(\tfrac{1}{2}mu^2)$     \;\;\;\;  &\;\;\;\;  $[(P_7/P_3)-(P_5/P_3)^2]^{1/2}$ \;\;\;\;&\;\;\;\;   0.143 \;\;\;\;&\;\;\;\;   0.0141 \\
$\triangle E_{BK}/\langle E_{BK}\rangle$  \;\;\;\;  &\;\;\;\;                       \;\;\;\;&\;\;\;\;   0.14 \;\;\;\;&\;\;\;\;    0.014 \\
$\langle E_R\rangle/(\tfrac{1}{2}\mu u_{12}^2)$  \;\;\;\;  &\;\;\;\;              Eq.(16)        \;\;\;\;&\;\;\;\;   0.0051 \;\;\;\;&\;\;\;\;  0.00020 \\
$\triangle E_R/(\tfrac{1}{2}\mu u_{12}^2)$  \;\;\;\;  &\;\;\;\;                   Eq.(17)        \;\;\;\;&\;\;\;\;   0.00697 \;\;\;\;&\;\;\;\; 0.000071 \\
$\triangle E_R/\langle E_R\rangle$  \;\;\;\;  &\;\;\;\;                             \;\;\;\;&\;\;\;\;   1.37 \;\;\;\;&\;\;\;\;   0.35 \\
\hline \hline
\end{tabular}
\end{minipage} \label{table2}
\end{table}

{\bf C. Rotating supersonic beams.}

For a supersonic beam from a rotating source, the velocity
distribution,
\begin{equation}
F(V) = V^2(V-V_{rot})\text{exp}\{ -[(V-w)/\triangle v]^2 \}
\end{equation}
involves the peripheral velocity of the source, $V_{rot}$, which
enters both in the $V - V_{rot}$ factor and in the flow velocity in
the laboratory frame, $w = u + V_{rot}$, with $u$ again relative to
the nozzle \cite{Gupta,Strebel}. In the slowing mode, when the rotor
spins contrary to the beam exit flow, $V_{rot} < 0$; in the speeding
mode, $V \ge V_{rot} > 0$. Also, in integrating over the velocity
distribution of Eq. (18), the lower limit depends on the sign of
$V_{rot}$.  In the slowing mode, the lower limit is small (typically
a few m$/$s) and represents a minimum, $V \ge V_{swat}$, necessary
to allow molecules to escape swatting by the rotor. As noted in the
Appendix, the swatting correction is extremely small, so we take
$V_{swat} = 0$. In the speeding mode, the lower limit may be large,
since $V \ge V_{rot}$ is required. Here, in addition to $x =
\triangle v/u$, it is convenient to specify two additional
variables: $y = w/u$ and $z = V_{rot}/u$. The Appendix gives
analytic results for the integrals, which are denoted by $P_n(x, y)$
for the slowing mode and $P_n(x, y, z)$ for the speeding mode.  The
corresponding velocity averages are given by
\begin{equation}
\langle(V/u)^k\rangle =  (P_{k+3} - zP_{k+2})/(P_3 - zP_2)
\end{equation}
where $P_n = P_n(x,y)$ or $P_n(x, y, z)$ for slowing or speeding ($z
< 0$ or $z > 0$), respectively.

For merged beam experiments with neutral atoms or molecules, pairing
a stationary supersonic source with a rotating source facilitates
adjusting the relative flow velocity. For a stationary source, to
adjust the flow velocity requires changing the temperature or, if
the beam species of interest is seeded in a carrier gas, changing
the seed ratio.   That is awkward and imprecise. For the rotating
source, the lab flow velocity, $w$, can be scanned easily and
precisely by merely changing the rotor speed.  The velocity width
for a beam from the rotating source is likely to be wider,
especially in the slowing mode (as $\triangle v/w > \triangle v/u$).
However, it will often be desirable to further narrow the velocity
spreads in both beams by means of electric or magnetic fields. Such
narrowing operations can benefit from having the peak intensities of
the beams preselcted to occur near the desired relative velocity.

The  merged beam  $\langle E_R\rangle$ and $\triangle E_R$ can be
obtained from Eqs.(16) and (17), using $\langle v_1^k\rangle$ from
Eq.(12) and $\langle v_2^k\rangle$ from Eq.(19). Figure 5 shows, for
both slowing ($y < 1$) and speeding modes ($y > 1$), how results
compare with those in Fig. 3 for a pair of stationary supersonic
beams (y = 1).    In panels (a) and (b), prominent minima occur
where the flow velocities match: $w_2 = yu_2 = u_1$, hence $u_2/u_1
= \text{tan}\phi = 1/y$.   In panel (c), the resolution ratio,
$\triangle E_R/\langle E_R\rangle$, is least good (largest values)
at the matching locations, but improves rapidly with modest
unmatching, dropping to near 0.1 within $\sim$15\% of the matching
peaks.

{\bf D. Distribution of relative kinetic energy.}

As seen in Figs. 3-5, to attain {\it both} low $\langle E_R\rangle$
and a small resolution ratio $\triangle E_R/\langle E_R\rangle$
requires that the merged beams have narrow velocity spreads. Even
for $x \sim 0.01$, to get a resolution ratio below $\sim$0.35
requires that the most probable velocities differ somewhat. To
examine further the competing aspects, we computed numerically the
distribution $P(E_R)$ of relative kinetic energy and its variation
with the velocity spread, extent of unmatching, and the merging
angle.

Figure 6 shows results for merged supersonic beams, both from
stationary sources; results with one beam from a rotating source are
similar. Panel (a) shows $P(E_R)$ distributions for beams having
velocity spreads of only 0.01 and merging angle $\theta = 1^\circ$
with ratios of flow velocities $u_2/u_1$ = 1, 0.95, 0.90, and 0.85.
Most striking is the form of $P(E_R)$ when the beam velocities are
both narrow and matched ($x = 0.01$, $u_2 = u_1$). Then the lower
limit of the relative kinetic energy, $E_R/E_R^\circ  < 1.5 \times
10^{-4}$, is sharply defined and $P(E_R)$ resembles qualitatively a
Poisson distribution. When the velocities become more and more
unmatched $P(E_R)$ becomes approximately Gaussian. In contrast, when
the beam velocities are broader (e.g., $x = 0.10$), the $P(E_R)$
distribution is roughly Gaussian for both matched and modestly
unmatched beam velocities.

Recasting Eq.(1) in the equivalent form,
\begin{equation}
E_R = \frac{1}{2}\mu\left[(V_1-V_2)^2 +
4V_1V_2\text{sin}^2(\theta/2)\right]
\end{equation}
makes evident why the lower limit for $E_R$ is sharply defined for
merged beams with narrow, matched velocities.  Then the first term
in $E_R$ becomes very small, yet as long as $\theta \not= 0$, the
second term will be appreciable because the parent beam velocities
are large. That second term will also be much less sensitive to
velocity spreads; in Fig. 6, panels (b) and (c) exhibit its role.

\section{EXPERIMENTAL PROSPECTS}

In study of chemical reactions at cold ($<$ 1 K) energies under
single-collision conditions, constraints enter that differ from
"warm" (typically $>$ 300 K) molecular beam experiments.  For
reactions without electronic or vibrational excitation, cold
reactive collisions cannot occur unless an activation barrier is
absent or exceptionally low or thin.   Such reactions are usually
exoergic.  The deBroglie wavelengths, long for the reactants, thus
become very short for the products: "waves in, particles out"
\cite{Herschbach}. Hence, disposal of energy and angular momentum
among product states is virtually the same as in the warm regime.
Also, as s-waves predominate in the entrance channel, the product
angular distribution for cold collisions is typically isotropic or
nearly so.   Accordingly, dynamical features observable in warm
collisions are not revealed in cold collisions.   In the cold
chemistry regime, the primary information to be sought is the
integral reaction cross section and its dependence on relative
kinetic energy or reactant excitation and/or interactions with
external fields.

Merged beams are well suited for measurements of total reaction
cross sections \cite{Phaneuf}. The advantage at very low collision
energies becomes immense as compared with experiments that slow both
reactants, because the centerline flux in a molecular beam (whether
effusive or supersonic) is proportional to the most probable
velocity in the beam.  Although merging imposes a very small angular
spread, that is compensated because the reactant beams meet in a
pencil-like volume, comparable in size to that in typical
crossed-beam experiments. Merged beams are also congenial for use
with pulsed sources, either stationary or rotating, which provide
much higher intensity than continuous sources.  Moreover, the option
of overlapping reactants or not in time enhances S/N discrimination.
Intensity is fostered too by the compactness of merged-beam geometry
which also somewhat simplifies incorporating auxiliary apparatus.

In order to facilitate obtaining estimates for any choice of
reactant atoms or molecules and velocity conditions, we have used
unitless ratios in Figs. 2-6 and Table II.   Here, in illustrative
discussion of experimental options, we revert to customary units:
energy in mK (or K); mass in amu, velocity in meters/sec.    The
merged beam technique can be used in many variants; we will consider
just two that serve to exemplify basic aspects.  We refer to these,
somewhat whimsically, as "ubiquitous" (ubiq) and "utopian" (utop),
to contrast what can be done using ordinary supersonic beams with
that requiring much more ambitious, sophisticated apparatus.

Table III displays the chief distinction: for ubiq, $x \sim 0.1$ is
typical, whereas for utop, $x \sim 0.01$ is as yet a challenging
goal. The corresponding ranges of averaged collision energy and rms
energy spread attainable are shown for either matched or 10\%
unmatched merged beams.  As the overall energy scale is governed by
$E_R^\circ = \tfrac{1}{2}\mu(u_1^2 + u_2^2)$, for either ubiq or
utop to obtain the lowest $\langle E_R\rangle$ and $\triangle E_R$
requires that the reduced mass and most probable beam velocities be
as small as feasible. Table III pertains to $\mu = 1$ amu, available
for reactions of an H atom with any considerably heavier reactant
partner.   The range of flow velocities considered, $u \sim 300$ to
600 m/s, extends as high as likely to be used.   Even for light
reactants, $u \sim 300$ m/s can be obtained by seeding in xenon
carrier gas.  That reduces intensity by a factor of typically $\sim
100$, but is routinely done as a precursor to most current methods
for slowing molecules.  Merged beams, especially if pulsed, can much
better afford such a drop in intensity because further slowing is
not required.

\begin{table}[H]
\begin{minipage}{16cm}
\centering \caption{Illustrative $\langle E_R\rangle$ and $\triangle
E_R$ values.$^a$}
\begin{tabular}{ccccccc}
\hline\hline
$u_1$ (m/s) \;\;&\;\; $u_2$ (m/s) \;\;&\;\;  $E_R^\circ$ (K)   \;\;& \multicolumn{2}{c}{\;\; $\langle E_R\rangle$ (mK)} &  \multicolumn{2}{c}{\;\; $\triangle E_R$ (mK)} \\
\;\;&\;\;     \;\;&\;\;\;\;\;     \;\;&\;\;   $x = 0.1$ \;\;&\;\;   $x = 0.01$ \;\;&\;\;   $x = 0.1$ \;\;&\;\;   $x = 0.01$ \\
\hline\let\thefootnote\relax\footnotetext{\;\;\;\;\;\;\;\;\;$^a$Kinetic
energies pertain to reduced mass $\mu$ = 1 amu and beam merging
angle $\theta = 1^\circ$, with velocity spreads $(x_1 = x_2)$ the
same in both beams. Entries derived from Table II, but rounded to one or two digits. Energy (mK) = 0.0605 mass(amu)$[\text{Velocity
(m/s)}]^2$. Corresponding deBroglie wavelength $\lambda_R(\text{nm})
= 395[\text{mass(amu)E(mK)}]^{-1/2}$.}
600   \;\;&\;\;  600   \;\;&\;\;   43 \;\;&\;\;   215 \;\;&\;\;   9 \;\;&\;\;   290  \;\;&\;\;   3\\
500   \;\;&\;\;  500   \;\;&\;\;   30 \;\;&\;\;   150 \;\;&\;\;   6 \;\;&\;\;   200  \;\;&\;\;   2\\
400   \;\;&\;\;  400   \;\;&\;\;   19 \;\;&\;\;   95 \;\;&\;\;    4 \;\;&\;\;   130  \;\;&\;\;   1.3\\
300   \;\;&\;\;  300   \;\;&\;\;   11 \;\;&\;\;   55 \;\;&\;\;    2 \;\;&\;\;    75  \;\;&\;\;   0.7\\

600   \;\;&\;\;  540   \;\;&\;\;   39 \;\;&\;\;   390 \;\;&\;\;  195 \;\;&\;\;   460  \;\;&\;\;   39\\
500   \;\;&\;\;  450   \;\;&\;\;   27 \;\;&\;\;   270 \;\;&\;\;  135 \;\;&\;\;   320  \;\;&\;\;   27\\
400   \;\;&\;\;  360   \;\;&\;\;   17 \;\;&\;\;   170 \;\;&\;\;   85 \;\;&\;\;   200  \;\;&\;\;   17\\
300   \;\;&\;\;  270   \;\;&\;\;   10 \;\;&\;\;   100 \;\;&\;\;   50 \;\;&\;\;   120  \;\;&\;\;   10\\
\hline \hline
\end{tabular}
\end{minipage} \label{table3}
\end{table}

As seen in Table III, the ubiq mode is capable of reaching $\langle
E_R\rangle$ as low as $\sim$ 55 mK for H atom reactions.  But, as
evident in Figs 3 and 5, the rms energy spread is larger, so the
resolution is miserable. In the cold collision regime, however, this
is not as severe a limitation as might be expected. According to
Wigner's threshold law \cite{Wigner}, for any exoergic inelastic or
reactive collision at sufficiently low energy the cross section
becomes proportional to $1/V_R$, so the rate coefficient becomes
independent of the relative kinetic energy.  State-of-the art
quantum scattering calculations \cite{Quemener} for a variety of
inelastic and reactive processes indicate the Wigner regime is
typically attained when $E_R$ drops below 100 mK ($< 10^{-5}$ eV),
which corresponds to a deBroglie wavelength of $\sim$ 40 nm or
longer for H atoms.

To achieve the utop mode by shrinking the velocity spreads to only
1\% of the most probable velocities, yet preserve adequate
intensity, the best means presently in prospect appears to be
deceleration by multistage Stark or Zeeman fields \cite{Hogan},
likely exploiting transverse focusing and phase stability together
with synchrotron-style storage rings \cite{Meeakker2}. Table III
presumes the resultant velocity distributions resemble Eq.(10).   If
so, the utop mode should be able to lower $\langle E_R\rangle$ to
$\sim$2 mK for H atom reactions, while reducing the energy spread to
0.7 mK, thereby improving the resolution to 35\%, a rather modest
level. As seen in Fig. 4 and Table II, in order to improve to the
resolution to 20\% would require that the beam velocities be
unmatched by $\sim$10\%, but that pushes  $\langle E_R\rangle$ up to
50 mK. As the Wigner regime is readily accessible for H atom
reactions, even for ubig, the elaborate experimental effort to
attain utop would not be justified. Obtaining decent resolution
would become much more worthwhile for reactions with considerably
larger reduced mass, however.  Then, even for utop, the Wigner
regime would lie mostly beyond the accessible range.

\section{DISCUSSION}

In summary, the merged-beam technique has several inviting virtues.
Foremost is the capability to study cold collisions with warm beams.
In contrast to methods that require slowing both collision partners,
the beam intensities can be much higher and the variety of molecular
species used much wider.   For collisions with small reduced mass,
especially H or D atoms with heavy molecules, relative kinetic
energy below 100 mK can be attained, using mostly standard, fairly
simple molecular beam apparatus.   However, to get low kinetic
energy for collisions with large reduced mass, and/or to obtain
decent energy resolution, will require making the beam velocity
spreads very small ($\sim$1\%), a challenging task which entails
developing far more elaborate apparatus.

The merged-beam experiments implemented in our laboratory aim both
to test the capabilities of ordinary, basic apparatus
\cite{Sheffield} and to explore in the cold regime H atom reactions
that have been well studied in the warm domain.  We note some
indicative aspects. Pulsed supersonic sources are operated at high
input pressures to enhance the familiar features of supersonic
beams: high intensity, narrowed velocity spreads, drastic cooling of
vibration and rotation.  One source is stationary, the other
rotating to enable readily adjusting its beam velocity to match that
from its sedate partner. The rotor source is suitable for any fairly
volatile molecule, while the stationary source can be used to
generate species that must be produced from precursors, such as
hydrogen, oxygen, or halogen atoms or free radicals. Initial
experiments use the H + NO$_2$ $\longrightarrow$ OH + NO reaction.
The H (or D) beam is supplied by dissociation of H$_2$ (or D$_2$) in
an RF discharge source of exceptional efficiency, evolved from a
design used in low-temperature NMR \cite{Boltnev}. The H beam,
seeded in Xe or Kr, has flow velocity in the range of 350 - 450 m/s
and spread $\sim$10\%. The rotating source provides the NO$_2$ beam,
currently with estimated velocity spread $\sim$20\%.  With the beams
merged at 420 m/s and $\theta = 1.5^\circ$, the predicted $\langle
E_R\rangle \sim 210$ mK and $\triangle E_R \sim 250$ mK. That is
well within the "cold" realm ($<$ 1 K), whereas the kinetic energy
is $\sim$11 K for the H beam and $\sim$490 K for the NO$_2$ beam.
The corresponding collisional deBroglie wavelength  $\lambda_R \sim
27$ nm.

Here we have considered only effusive and supersonic beams, the most
familiar.   There is, however, a growing repertoire of other types,
developed with focus on slowing.  Some may offer properties
advantageous for merged-beams, which shift the focus to narrowing
velocity spreads.   In particular, we mention the recently developed
cryogenically cooled buffer gas beams \cite{Hutzler}, which operate
in the intermediate regime between effusive and supersonic flow.
There the Knudsen number (essentially the ratio of Reynolds number
and Mach number) is $Kn \sim 1 - 10^{-2}$, whereas for effusive
beams $Kn > 1$ and for supersonic beams typically $Kn \le 10^{-3}$.
In a prototype case producing a ThO beam \cite{Hutzler2}, a
continuous gas flow maintains a stagnation density of Ne buffer gas
in thermal equilibrium within a cell cooled to 18 K.   A solid
ThO$_2$ target is mounted to the cell wall.  A YAG laser pulse
vaporizes part of the target and ejects ThO molecules that are
cooled by collision with the Ne buffer as they flow out of the cell.
A key effect is "hydrodynamic entrainment" that occurs when the time
for the ThO molecules to exit the cell is less than the diffusion
time to the cell walls.  Remarkably, the molecular beam has
intensity exceeding that for a supersonic beam, very low internal
temperature, and its most probable velocity, governed by the cooling
Ne, is $<$ 200 m/s. The velocity spread is broad, but the high
intensity and low speed would permit much narrowing via velocity
selection.   Such cryogenically cooled and buffered beams seem
likely to become widely applicable, perhaps along with merged-beams.

\section*{ACKNOWLEDGEMENTS}
We are grateful for support of this work by the National Science
Foundation (under grant CHE-0809651), the Office of Naval Research
and the Robert A. Welch Foundation (under grant A-1688).  We thank
Ronald Phaneuf of the University of Nevada for helpful
correspondence about his work with merged high-energy beams.

\renewcommand{\theequation}{A\arabic{equation}}
\setcounter{equation}{0}  
\section*{APPENDIX: INTEGRALS FOR VELOCITY AVERAGES}

The averages, $\langle V^k\rangle$, over the velocity flux
distributions specified in Eqs. (10) and (18) for supersonic beam
sources involve two related integrals
\begin{eqnarray}
P_n(x,y) & = & \int_0^\infty t^n
\text{exp}\left[-\left(\frac{t-y}{x}\right)^2\right]dt \\
& = & \frac{\sqrt{\pi}}{16}\left[1+\text{Erf}(y/x)\right]A_n(x,y) +
\frac{x^2}{8}\text{exp}\left[-(y/x)^2\right]B_n(x,y)
\end{eqnarray}
\begin{eqnarray}
P_n(x,y,z) & = & \int_z^\infty t^n
\text{exp}\left[-\left(\frac{t-y}{x}\right)^2\right]dt \\
& = & \frac{\sqrt{\pi}}{16}\left[1+\text{Erf}(1/x)\right]A_n(x,y) +
\frac{x^2}{8}\text{exp}\left[-(1/x)^2\right]
\left[B_n(x,y)+C_n(x,y,z)\right]
\end{eqnarray}
where, $n=k+3$, Erf is the error function, and $x=\triangle v/u$,
$y=w/u$, $z=V_{rot}/u=y-1$. For a stationary source, only $P_n(x,
1)$ enters, as $y = 1$ and $z = 0$; for a rotating source, $P_n(x,
y)$ pertains to the slowing mode, with $0 < y < 1$ and $z < 0$, and
$P_n(x, y, z)$ to the speeding mode, with $y > 1$ and $z > 0$.
Analytical formulas for the $A_n(x, y)$, $B_n(x, y)$ and $C_n(x, y,
z)$ functions can be obtained by evaluating the $P_n$ in successive
steps using integration by parts. Table I of the text gives, for $n$
= 2 to 7, explicit expressions for $A_n(x, y)$; Table IV below gives
$B_n(x, y)$ and $C_n(x, y, z)$.

In the range of interest here, $x < 0.3$ and $y > 0.5$, the first
term in Eqs. (A2) and (A4) is much larger than the others, and the
error function is very close to unity. Accordingly, good
approximations for $P_n(x, y)$ and $P_n(x, y, z)$ can be obtained by
dropping the exponential part and replacing the Erf function by
unity for compensation; thus,
\begin{eqnarray}
P_n(x,y) \approx P_n(x,y,z) \approx \frac{\sqrt{\pi}}{8}A_n(x,y) =
\sqrt{\pi}\sum_{s=0}^{[n/2]} \left( \begin{array}{c} n \\ 2s
\end{array} \right) \frac{(2s-1)!!}{2^s}x^{2s+1}y^{n-2s}
\end{eqnarray}
For $x<0.3$ and $y>0.5$, the error is $<0.03\%$. We note that the
swatting correction mentioned under Eq.(18) of the text may be
obtained from $\triangle P_n = P_n(x, y) - P_n(x, y, z*)$ with $z* =
V_{swat}/u$. However, in the range of interest,  $\triangle P_n$ is
extremely small, much less than 0.03\% so taking $V_{swat} = 0$ is
quite justified.

The  velocity averages are given by
\begin{equation}
\langle (V/u)^k\rangle = (P_{k+3} - zP_{k+2})/(P_3 - zP_2)
\end{equation}
a generic formula that includes Eqs.(5), (12), and (19) of the text.
For a stationary supersonic source, $P_n = P_n(x) = P_n(x, 1)$; for
a rotating source, $P_n = P_n(x,y)$ or $P_n(x, y, z)$ for slowing or
speeding, respectively.   The effusive beam case corresponds to $x =
\alpha$, $y = 0$, $z = 0$, with $P_n(x, 0) =
\tfrac{1}{2}\alpha^{n+1}\Gamma[(n + 1)/2]$. Results for averages
over number density distributions rather than flux distributions can
be obtained merely by setting $n = k + 2$.

\begin{table}[H]
\caption{Formulas for $B_n(x,y)$ and $C_n(x,y,z)$.}
\begin{tabular}{ccc}
\hline\hline
$n$ \;\;\;\;&\;\;\;\; $B_n(x,y)$ \;\;\;\;&\;\;\;\; $C_n(x,y,z)$   \\
\hline
2 \;\;\;\;&\;\;\;\;   $4y$  \;\;\;\;&\;\;\;\;   $4z$ \\
3 \;\;\;\;&\;\;\;\;   $4(y^2+x^2)$   \;\;\;\;&\;\;\;\;  $4z(z+y)$ \\
4 \;\;\;\;&\;\;\;\;   $2(2y^2+5x^2)y$   \;\;\;\;&\;\;\;\;  $2[2z^3+2yz^2+(2y^2+3x^2)z]$   \\
5 \;\;\;\;&\;\;\;\;   $2(2y^4+9x^2y^2+4x^4)$   \;\;\;\;&\;\;\;\;  $2[2z^4+2yz^3+2(y^2+x^2)z^2+(2y^3+7x^2y)z]$   \\
  &                          \;\;\;\;&\;\;\;\; $4z^5+4yz^4+2(5x^2+2y^2)z^3+2(2y^3+9x^2y)z^2$ \\[-1.8ex]
\raisebox{1.5ex}{6}\;\;\;\;&\;\;\;\;\raisebox{1.5ex}{$(4y^4+28x^2y^2+33x^4)y$}
\;\;\;\;&\;\;\;\; +$(4y^4+24x^2y^2+15x^4)z$
\\[1.8ex]
  &                          \;\;\;\;&\;\;\;\; $4z^6+4yz^5+4(3x^2+y^2)z^4+2(2y^3+11x^2y)z^3$ \\[-1.8ex]
\raisebox{1.5ex}{7}\;\;\;\;&\;\;\;\;\raisebox{1.5ex}{$4y^6+40x^2y^4+87x^4y^2+24x^6$}
\;\;\;\;&\;\;\;\;
+$2(2y^4+15x^2y^2+12x^4)z^2+(4y^5+36x^2y^3+57x^4y)z$
\\[1.8ex]
\hline\hline
\end{tabular}
\label{table-BC}
\end{table}

\newpage

\begin{figure}[htp]
\begin{center}
\includegraphics[width=0.5\textheight]{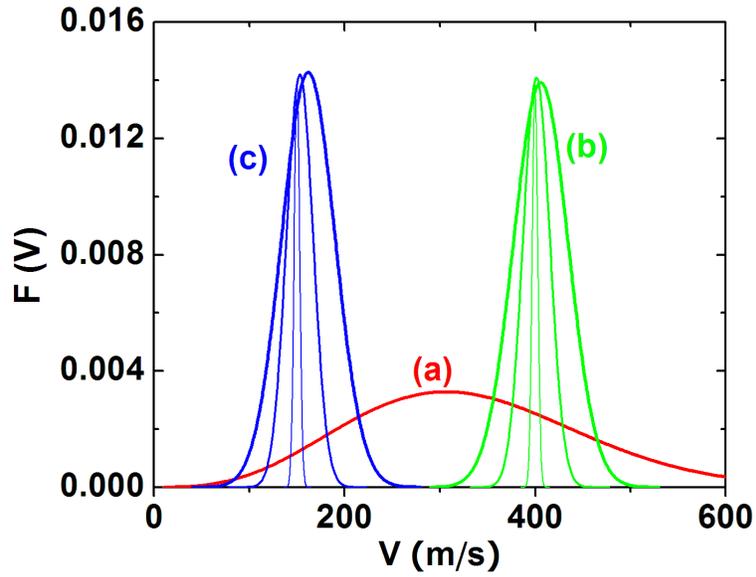}
\end{center}
\caption{(Color online) Velocity distributions of molecular flux,
$F(V)$, for beams formed (a) by effusive flow, (b) supersonic
expansion from a stationary source, and (c) a rotating supersonic
source, defined by Eqs.(4), (10), and (18), respectively. Parameters
for (a) are $T_0 = 300 K$; $\alpha = 250$ m/s; for (b) and (c) flow
velocities are $u = 400$ m/s and $w = u + V_{rot} = 250$ m/s and
widths $\triangle v/u$ = 0.01, 0.05, 0.10. For (b) and (c) the
widths also influence somewhat the most probable velocity.}
\label{Figure1}
\end{figure}

\newpage

\begin{figure}[htp]
\begin{center}
\includegraphics[width=0.4\textheight]{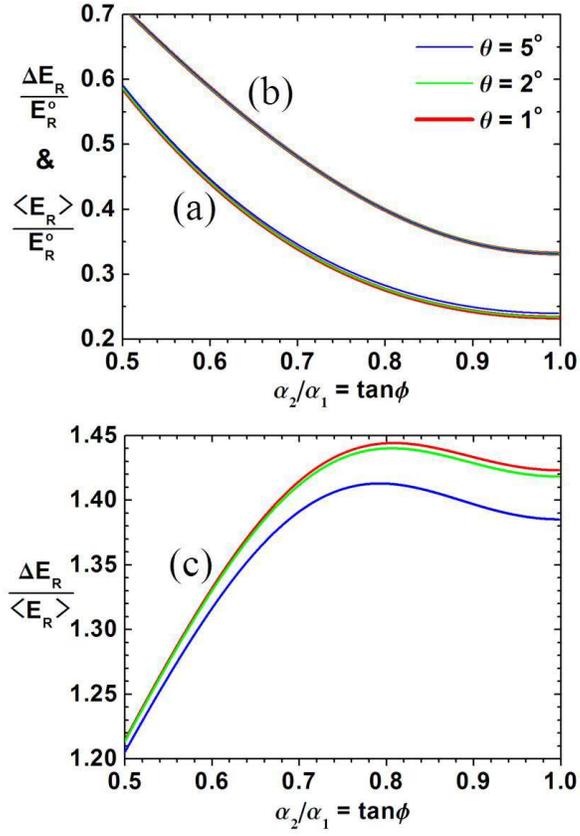}
\end{center}
\caption{(Color online) Properties of merged effusive beams, from
Eqs.(8) and (9), obtained by averaging over beam velocity
distributions. (a) Averaged relative kinetic energy; (b) its rms
spread; (c) ratio of spread to averaged kinetic energy, defining the
available resolution. Ordinate energy scale is $E_R^\circ =
\frac{1}{2}\mu(\alpha_1^2+\alpha_2^2)$; abscissa scale pertains to
ratio $\alpha_2/\alpha_1$ ranging from $\alpha_2 =
\frac{1}{2}\alpha_1$ to $\alpha_2 = \alpha_1$. Results are shown for
intersection angles of $\theta = 1^\circ$, $2^\circ$ and $5^\circ$.}
\label{Figure2}
\end{figure}

\newpage

\begin{figure}[htp]
\begin{center}
\includegraphics[width=0.4\textheight]{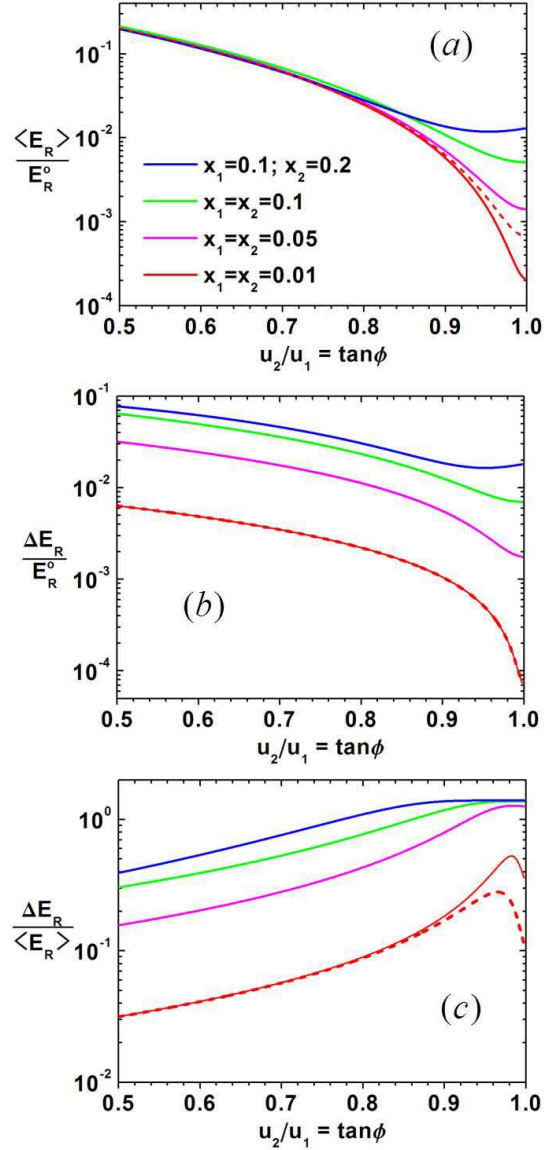}
\end{center}
\caption{(Color online) Properties of merged supersonic beams, from
Eqs.(16) and (17), obtained by averaging over beam velocity
distributions. (a) Averaged relative kinetic energy; (b) its rms
spread; (c) ratio of spread to averaged kinetic energy. For (a) and
(b) the ordinate energy scale is $E_R^\circ = \frac{1}{2}\mu(u_1^2 +
u_2^2)$. The abscissa scale pertains to the ratio $u_2/u_1$, which
ranges from $u_2 = \frac{1}{2}u_1$ to $u_2 = u_1$.  Results are
shown for an intersection angle of $\theta = 1^\circ$ and four sets
of velocity spreads:  $x_1 = x_2$ = 0.01; 0.05; 0.1; and $x_1=0.1$,
$x_2=0.2$. Dashed curves included for the $x = 0.01$ case are for
$\theta = 2^\circ$.} \label{Figure3}
\end{figure}

\newpage

\begin{figure}[htp]
\begin{center}
\includegraphics[width=0.5\textheight]{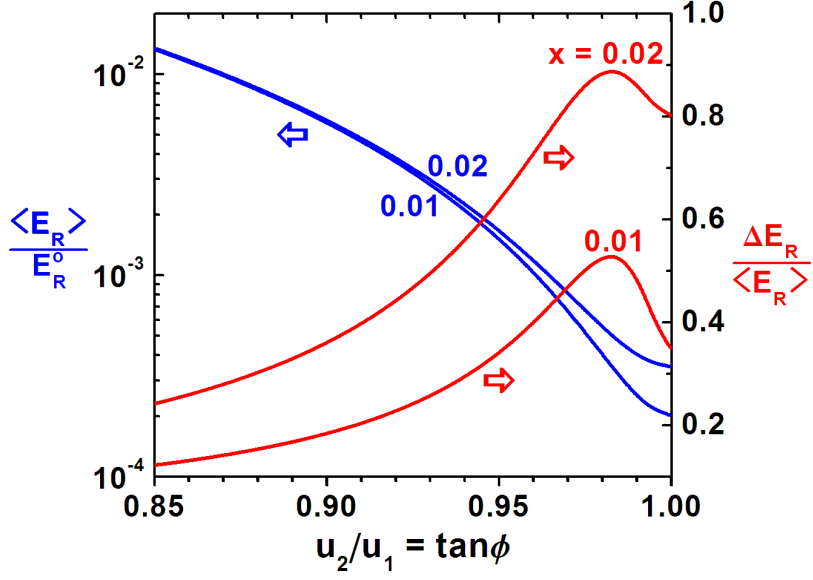}
\end{center}
\caption{(Color online) Variation with extent of matching of merged
beams flow velocities, $u_2/u_1$, for relative kinetic energy,
$\langle E_R\rangle/E_R^\circ$ (left ordinate scale) and resolution
ratio, $\triangle E_R/\langle E_R\rangle$ (right ordinate scale).
Curves pertain to velocity spreads in beams of $x_1 = x_2 = 0.01$
and 0.02 and merging angle of $\theta = 1^\circ$.} \label{Figure4}
\end{figure}

\newpage

\begin{figure}[htp]
\begin{center}
\includegraphics[width=0.4\textheight]{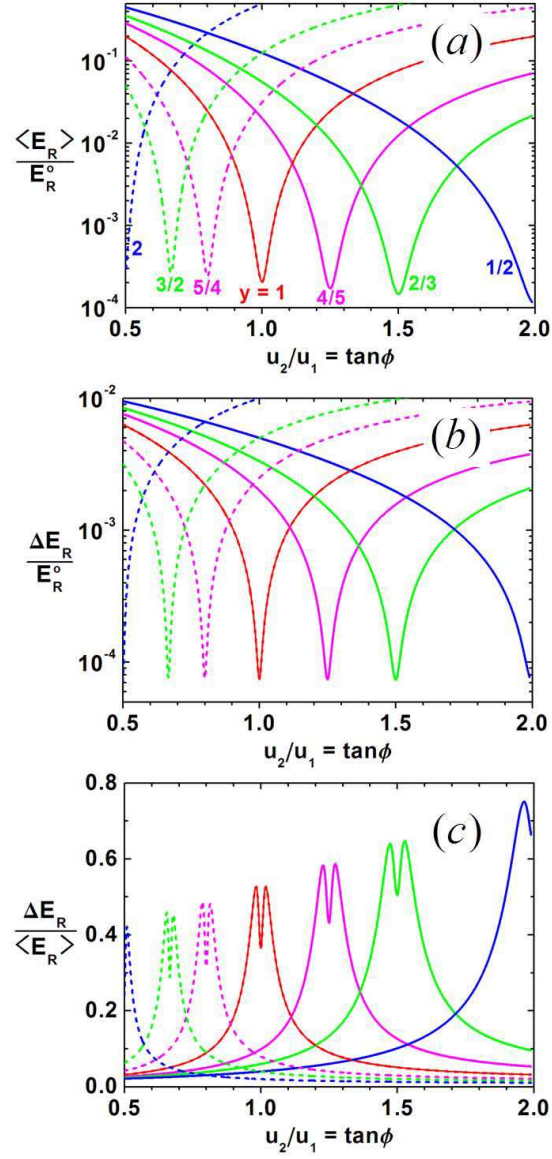}
\end{center}
\caption{(Color online) Properties of merged supersonic beams, one
from a stationary source, the other from a rotating source.   As in
Fig. 3, (a) shows the averaged relative kinetic energy; (b) its rms
spread; (c) the resolution ratio, spread to averaged kinetic energy.
Results shown are for $\theta = 1^\circ$ and $x = 0.01$. Both (a)
and (b) exhibit pronounced minima where the matching condition
holds: $w_2 = u_1$ and hence $u_2/u_1 = 1/y$. Full curves show
results for slowing mode, with $y = 4/5$, $2/3$, and $1/2$; dashed
curves are for speeding mode, with $y = 5/4$, $3/2$, and 2.}
\label{Figure5}
\end{figure}

\newpage

\begin{figure}[htp]
\begin{center}
\includegraphics[width=0.4\textheight]{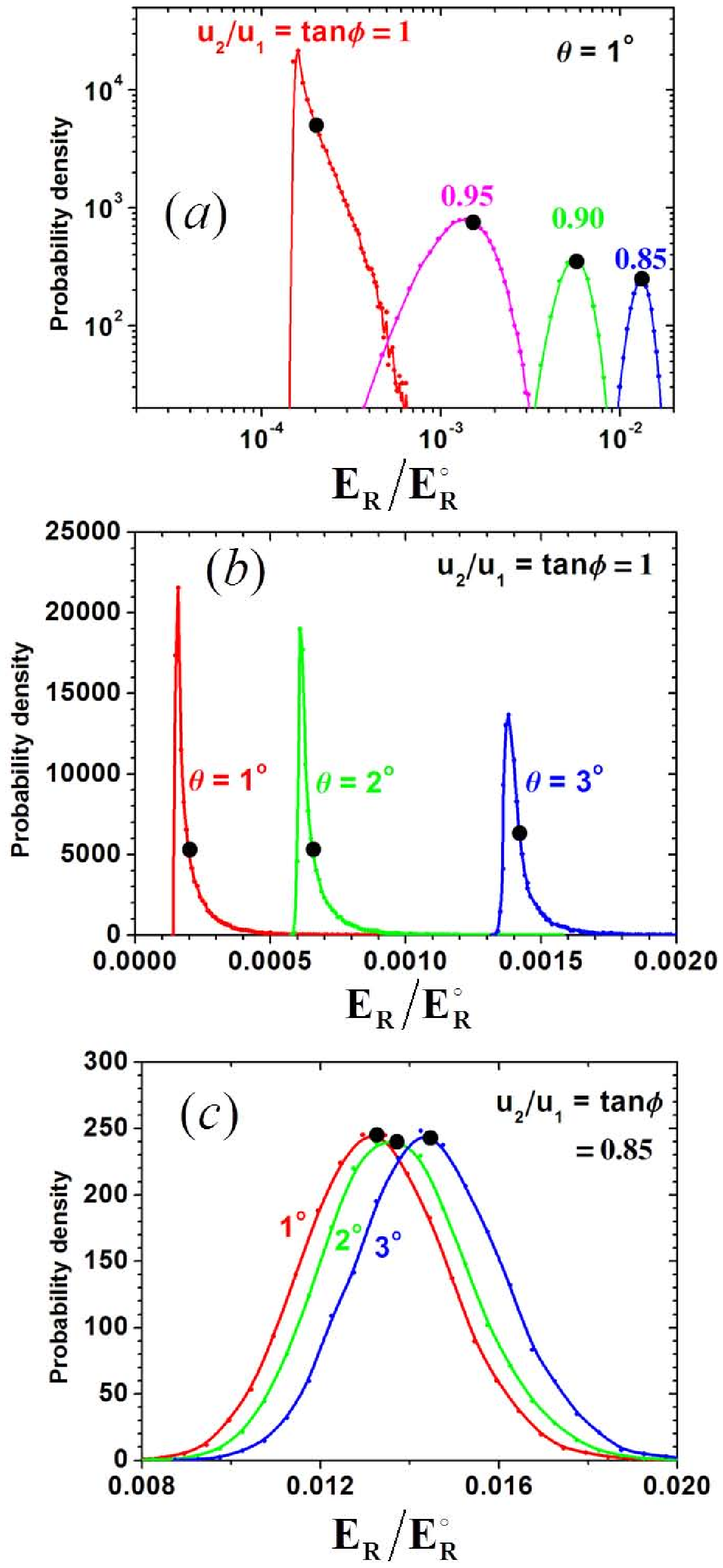}
\end{center}
\caption{(Color online) Distributions of the relative kinetic
energy, $P(E_R)$, for merged supersonic beams from stationary
sources. Values of $\langle E_R\rangle$ are indicated by black dots.
Abscissa scale is in units of $E_R^\circ = \frac{1}{2}\mu\left(u_1^2
+ u_2^2\right)$, as used in Figs.3-5. (a) For merging angle
$\theta=1^\circ$ and various ratios of the flow velocities, $u_2/u_1
= 1$ to 0.85 and velocity widths $x_1 = x_2 = 0.01$. Note log-log
plot is used.  (b) For matched flow velocities, $u_2/u_1 = 1$ and $x
= 0.01$, but merging angle varied to illustrate its role in Eq.(20).
(c) For flow velocities unmatched by $\sim$15\% ($u_1u_2 = 0.85$)
and $x = 0.01$ with $\theta = 1^\circ$, $2^\circ$, $3^\circ$, to
illustrate the reduced role of the merging angle when the first term
in Eq.(20) becomes predominant.} \label{Figure6}
\end{figure}


\begin{thebibliography}{71}

\bibitem{Book1} {\it Cold Molecules: Theory, Experiment, Applications}, R. V. Krems, W. C. Stwalley, and B. Friedrich, Eds., (Taylor and Francis, London, 2009).

\bibitem{Carr} {\it Focus on Cold and Ultracold Molecules}, L. D. Carr and J. Ye, Eds., New J. Phys. {\bf 11}, 055009 (2009).

\bibitem{Faraday} {\it Cold and Ultracold Molecules}, Faraday Disc. {\bf 142}, (Royal Soc. Chem., London, 2009).

\bibitem{Dulieu} {\it Physics and Chemistry of Cold Molecules}, Phys. Chem. Chem. Phys. {\bf 13}, 18703, O. Dulieu, R. Krems, M. Weidemuller, and S. Willitsch, Eds. (2011).

\bibitem{Ospelkaus} S. Ospelkaus, K.-K. Ni, D. Wang, M. H. G. de Miranda, B. Neyenhuis, G. Quemener, P. S. Julienne, J. L. Bohn, D.S. Jin, J. Ye, Science {\bf 327}, 853 (2010).

\bibitem{Miranda} M. H. G. de Miranda, A. Chotia, B. Neyenhis, D. Wang, G. Quemener, S. Ospelkaus, J. L. Bohn, J. Ye, D. S. Jin, Nature Physics {\bf 7}, 502 (2011).

\bibitem{Meerakker} S. Y. T. van de Meerakker, H. L. Bethlem, and G. Meijer, Nature Physics {\bf 4}, 595 (2008).

\bibitem{Bell} M. T. Bell and T. P. Softley, Mol. Phys. {\bf 107}, 99 (2009).

\bibitem{Hogan} S. D. Hogan, M. Motsch, and F. Merkt, Phys. Chem. Chem. Phys. {\bf 13}, 18705 (2011).

\bibitem{Sheffield} L. Sheffield, M. Hickey, V. Krasovitsky, K. D. D. Rathnayaka, I. F. Lyuksyutov and D. R. Herschbach, Rev. Sci Instrum. (in press 2012).

\bibitem{Phaneuf} R. A. Phaneuf, C. C. Havener, G. H. Dunn, and A. Muller, Rep. Prog. Phys. {\bf 62}, 1143 (1999).

\bibitem{Book2000} H. Pauly, in \textit{Atom, Molecule, and Cluster Beams}, H. Pauly, Ed. (Springer,
2000), p. 8.

\bibitem{Gupta} M. Gupta and D. Herschbach, J. Phys. Chem. A {\bf 105}, 1626 (2001).

\bibitem{Meeakker2} S. Y. T. van de Meeakker and G. Meijer, Faraday Disc. {\bf 142}, 113 (2009).

\bibitem{Book1988} D. R. Miller, in \textit{Atomic and Molecular Beam Methods}, Vol. I, G. Scholes, Ed. (Oxford Univ. Press, New York, 1988), p. 14.

\bibitem{Klots} C. Klots, J. Chem. Phys. {\bf 72}, 192 (1980).

\bibitem{Strebel} M. Strebel, F. Stienkemeier, and M. Mudrich, Phys. Rev. A {\bf 81}, 033409 (2010).

\bibitem{Herschbach} D. Herschbach, Faraday Disc. {\bf 142}, 9 (2009).

\bibitem{Wigner} E. P. Wigner, Phys. Rev. {\bf 73}, 1002 (1948).

\bibitem{Quemener} G. Quemener, N. Balakrishnan and A. Dalgarno, {\it Inelastic collisions and chemical reactions of molecules at ultracold temperatures,
in Cold Molecules: Theory, Experiment, Applications},  R. V. Krems,
W. C. Stwalley, and B. Friedrich, Eds., (Taylor and Francis, London,
2009), p. 69.

\bibitem{Boltnev} R. E. Boltnev, V. V. Khmelenko, and D. M. Lee, Low Temp. Phys. {\bf 36}, 382 (2010).

\bibitem{Hutzler} N. R. Hutzler, H.-I Lu, and J. M. Doyle, Chemical Reviews (in press).

\bibitem{Hutzler2} N. R. Hutzler, M. F. Parsons, Y. V. Gurevich, P. W. Hess, E. Petrik, B. Spaun, A. C. Vutha, D. DeMille, G. Gabrielse, J. M. Doyle, Phys. Chem. Chem. Phys. {\bf 13}, 18976 (2011).

\end{thebibliography}
\end{document}